# An Empirical Study of Obsolete Answers on Stack Overflow

Haoxiang Zhang, Shaowei Wang, Tse-Hsun (Peter) Chen, Ying Zou, *Senior Member, IEEE,* and Ahmed E. Hassan, *Fellow, IEEE*

**Abstract**—Stack Overflow accumulates an enormous amount of software engineering knowledge. However, as time passes, certain knowledge in answers may become obsolete. Such obsolete answers, if not identified or documented clearly, may mislead answer seekers and cause unexpected problems (e.g., using an out-dated security protocol). In this paper, we investigate how the knowledge in answers becomes obsolete and identify the characteristics of such obsolete answers. We find that: 1) More than half of the obsolete answers (58.4%) were probably already obsolete when they were first posted. 2) When an obsolete answer is observed, only a small proportion (20.5%) of such answers are ever updated. 3) Answers to questions in certain tags (e.g., node.js, ajax, android, and objective-c) are more likely to become obsolete. Our findings suggest that Stack Overflow should develop mechanisms to encourage the whole community to maintain answers (to avoid obsolete answers) and answer seekers are encouraged to carefully go through all information (e.g., comments) in answer threads.

✦

## 1 INTRODUCTION

TECHNICAL Q&A websites are becoming an important and popular platform for knowledge sharing and learning. They have revolutionized how users seek knowledge on the Internet. When users face unsolvable problems, they often try to search for solutions via search engines (e.g., Google). A case study shows that Google developers perform an average of 12 code search queries each weekday [1]. Search engines commonly direct users to technical Q&A websites in response to their queries. As a prominent example, Stack Overflow, one of the most popular technical Q&A websites, has collected an enormous amount of knowledge, which includes 15 million questions, 23 million answers, and 62 million comments as of September 2017[1].

Software systems evolve at a rapid pace nowadays. For instance, Android has released 16 major versions and 53 minor versions since September 2008 (as of August 2018) [2]. Android is evolving at a rate of 115 API updates per month on average according to a study by McDonnell et al. [3]. Such rapid evolution may make the knowledge in some Stack Overflow answers obsolete over time. Fig. 1 presents an example of such a case, where the user was directed from Google to a Stack Overflow answer. However, the user found that the content of the answer thread (including the answer and the discussions in the comments) was obsolete and asked whether Stack Overflow has any mechanisms

- H. Zhang, S. Wang and A. E. Hassan are with the Software Analysis and Intelligence Lab (SAIL), Queen's University, Kingston, Ontario, Canada. E-mail: hzhang,shaowei,ahmed@cs.queensu.ca
- T. Chen is with the Department of Computer Science and Software Engineering, Concordia University, Montreal, Quebec, Canada. E-mail: peterc@encs.concordia.ca
- Y. Zou is with the Department of Electrical and Computer Engineering, Queen's University, Kingston, Ontario, Canada. E-mail: ying.zou@queensu.ca
- Shaowei Wang is the corresponding author.

1. https://data.stackexchange.com/stackoverflow/

to handle such a situation[2]. Additionally, a survey of 453 Stack Overflow users reports that outdated code on Stack Overflow is one of the most important issues that users complain about [4].

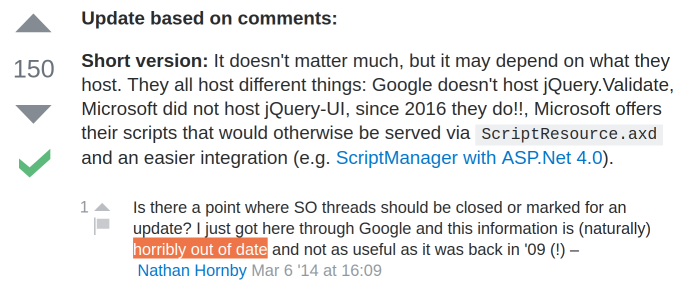

Fig. 1: An example of a user complaining in a comment that the Stack Overflow answer thread (including the answer and the discussions in the comments) is obsolete.

Obsolete answers are detrimental to answer seekers. For example, a user found a piece of code that matches his/her needs and reused it in his/her own project. However, the user may not realize that the used APIs in the code are obsolete. Using such obsolete APIs could potentially result in software quality problems (e.g., using an outdated security framework API), and may increase maintenance difficulties. Therefore, it is necessary to provide insights on how to track or alleviate this problem.

In this paper, we study 52,177 Stack Overflow answer threads (each answer thread includes all answers to a question (i.e., accepted & not-accepted answers) and all the comments that are associated with them) to understand how the knowledge that is embedded in answer threads becomes obsolete and the characteristics of such obsolete

2. https://stackoverflow.com/posts/comments/33754357/

answers, and to provide actionable suggestions. We also perform a qualitative study to understand the evidence that users provide to support their obsolete observations and the activities that users perform after an answer is observed as obsolete. We structure our study by answering the following research questions:

- **RQ1: What happens when an answer is observed as obsolete?**
  More than half of the studied obsolescence observations refer to answers that were probably already obsolete when they were first posted. Most users did not update obsolete answers or add new answers to address the observed obsolescence. On average, it took 118 days for users to react to an observed obsolete answer.
- **RQ2: Are answers to questions with particular tags more likely to become obsolete than other answers?**
  Answers to questions that are associated with certain tags (e.g., node.js, ajax, android, and objective-c) are more likely to become obsolete.
- **RQ3: What are the potential reasons for answers to become obsolete?**
  The majority of the answers become obsolete due to the evolution of their associated programming languages and/or third party libraries, APIs, and frameworks. Therefore, users need to pay more attention to such answers when looking for answers on Stack Overflow.
- **RQ4: Who observes obsolete answers and what evidence do they provide?**
  The majority of the obsolete answers were not observed by the original answerers. Also, most obsolescence observations are supported by evidence (e.g., updated information, a version information, or a reference).

Based on our observations, we provide actionable suggestions for Stack Overflow to alleviate the problem of obsolete answers. For example, an automated tool based on machine learning techniques or even simple keyword search could be built to identify existing obsolete answers on Stack Overflow, or help answerers identify obsolete answers in real-time as an answer is being typed. Moreover, Stack Overflow should develop mechanisms (e.g., rewarding badges or reputation scores) to encourage the whole community to maintain answers and flag obsolete answers. We also provide suggestions for users. For example, answerers are encouraged to include whenever possible information about the valid version or time of the knowledge when contributing answers. Answer seekers are encouraged to carefully go through the comments that are associated with answers in case the obsolescence of an answer is noted in the comments, especially for the answers in questions that are related to particular tags (e.g., node.js, ajax, android, and objective-c). We also shared our findings with Stack Overflow developers who concurred with our findings, and they were interested in investigating approaches to generate version tags to indicate the valid version for a platform or programming language used in obsolete answers.

**Paper Organization:** The rest of the paper is organized as follows. Section 2 presents the background. Section 3 introduces our data collection process. Section 4 presents the results of our research questions. Section 5 discusses the implications of our study. Section 6 presents the potential threats to the validity of our observations. Section 7 discusses related work. Finally, Section 8 concludes the paper.

## 2 BACKGROUND

In this section, we briefly introduce the mechanism of question answering on Stack Overflow and discuss how answers on Stack Overflow can become obsolete.

### 2.1 The Question Answering Mechanism on Stack Overflow

Stack Overflow provides a platform for the asking and answering of questions. Askers post questions which include a textual description on Stack Overflow. Askers can include code snippets and other references (e.g., URLs or images) to enrich their posted question. Each question may receive multiple answers from different answerers. However, at most one answer could be accepted by the asker as the *accepted answer* to indicate that this particular answer is the most suitable/correct one.

In the rest of the paper, we refer to a question, its corresponding answers (i.e., both accepted and not-accepted answers) and all the associated comments with these answers together as a *question thread*. We refer to an answer (could either be accepted or not-accepted answers) and its comments as an *answer thread*.

Users tag questions[3] into well-defined categories. Tags capture the topics with which a question is associated. Each question can have at most five tags and must have at least one tag. Askers need to specify the tags when they create a question. In the rest of the paper, we say that an answer is associated with a particular tag if the answer belongs to a question that is associated with that tag. In RQ2, we study whether answers to questions that are associated with particular tags are more likely to become obsolete.

### 2.2 Obsolete Answers on Stack Overflow

As we noted in Section 1, Stack Overflow users complain about the obsolescence of answers. There are various reasons that an answer could become obsolete on Stack Overflow. For instance, APIs could become deprecated later on when a new API version is released. For a better understanding of answer obsolescence on Stack Overflow, we present the possible activities that could happen after an answer becomes obsolete in Fig. 2. An answer probably becomes obsolete after some time since its creation (alternatively an answer might be obsolete even as it is being posted) (see Section 4.1). An obsolete answer probably would be observed by a user on Stack Overflow (i.e., obsolescence observation). Users may also discuss the obsolescence afterwards and update their answers accordingly.

Obsolete answers are problematic on Stack Overflow. However, there exists no mechanisms in place today to alleviate the problem of obsolete answers. Thus, in this paper, we wish to closely examine the obsolescence of answers in an effort to propose ways to help Stack Overflow deal with such answers in an effective & efficient manner. To do so, we investigate what happens once someone identifies that an answer has become obsolete and whether answers in questions that are associated with particular tags are more likely to become obsolete. We also investigate who observes obsolete answers and what evidence do they provide to support their observations.

---

3. https://stackoverflow.com/help/tagging

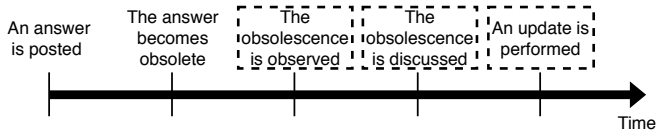

Fig. 2: A possible flow of activities that could occur after an answer becomes obsolete. Activities in dotted box are optional and might not happen in all cases.

Based on our study, obsolete answers could be categorized into two classes: *legacy* or *invalid*. We consider an obsolete answer as a *legacy* answer if it can still be used or applied, but it may not be recommended anymore since a newer answer might be better or more appropriate. For example, a comment[4] points out that an answer is "obsolete in Rails >= 3.0.0", which indicates that the accepted answer only applies to Rails version 3.0.0 or below. Nevertheless, users who use earlier versions may find this answer still useful. On the other hand, an *invalid* answer indicates that the obsolete answer is not valid or that it no longer works. Users who might have successfully applied the particular answer earlier would now run into errors or complete failures. One example of an invalid answer is related to an old http protocol (such as RFC 2616[5]), which is deprecated. For example, a comment[6] mentions that "RFC 2616 has been obsoleted".

Thus, we are interested in investigating obsolete answers on Stack Overflow, to understand obsolescence reasons that happen and to provide some insights into addressing the obsolescence of answers.

## 3 DATA COLLECTION

In this section, we describe how we collect the dataset that we used to answer our research questions.

To understand the obsolescence of answers on Stack Overflow, we need to identify answer threads (both accepted and not-accepted answers) with obsolete knowledge. As we introduce in Section 2, users occasionally leave comments to indicate that an answer is obsolete (see Fig. 1). Based on this observation, we identify answer threads that have obsolete knowledge using both of the two following criteria:

1) A comment in an answer thread contains one of the keywords "deprecated", "outdated", "obsolete" or "out of date".
2) The same keyword from criteria 1 ("deprecated", "outdated", "obsolete" or "out of date") does not

---
[4]. https://stackoverflow.com/posts/comments/30559321/
[5]. https://www.ietf.org/rfc/rfc2616.txt
[6]. https://stackoverflow.com/posts/comments/61676900/

appear in the question (including the question title and body) of its thread or any of its answers. The reason behind this criteria is that if the keyword appears in the content of a question or an answer, it may indicate that the question or answer itself is related to an "obsolete" topic rather than being a sign that the answer is likely obsolete.

The purpose of our selection criteria is not to collect all possible answer threads with obsolete knowledge, but to collect sufficient data for a relatively comprehensive analysis, while minimizing the bias that is caused by false positives.

We downloaded the Stack Overflow data from archive.org[7]. The data was published on August 31, 2017 by the Stack Exchange community. The data contains information about badges, comments, post history, post links, posts, tags, users, and votes. Using our selection criteria, we ended up with 52,177 answer threads, which include 58,201 comments that mention obsolescence. These collected threads span 12,629 tags. We published our data set including the labeled data online[8].

The accuracy of our heuristic-based approach is 75% based on our manual verification of a statistically representative sample with a 99% confidence level and a 5% confidence interval. For each observed answer obsolescence, we examine the support evidence from the user who observed the obsolescence together with online information (e.g., documentation for API, programming language, and framework), to verify if the answer is really obsolete. If no obsolescence is identified, we label it as a false positive. 167 answers out of the 669 are false positives. The two main reasons for the false positive cases are: 1) Instead of indicating the obsolescence of an answer in the comment, the content that is discussed by users in the comments is related to certain topics which use our keywords of interest (e.g., "obsolete" and "out of date"). For example, in a comment[9] the user mentions that "unless you have some kind of locking mechanism (which I'd argue against), the result of the call would be obsolete as soon as you got it". This comment did not indicate the obsolescence of the answer. 2) Users either ask whether the answer is obsolete or express that the answer probably will become obsolete soon. For example, in a comment[10] "because php is changing a lot and in upcoming versions this might be deprecated", the user did not observe any specific obsolete software artifact in the answer, but just simply expressed the user's general feeling that PHP is evolving very fast and that it is obsolete-prone.

## 4 CASE STUDY RESULTS

### 4.1 RQ1: What happens when an answer is observed as obsolete?

**Motivation:** It is very important to keep answers up-to-date on Stack Overflow as we noted in Section 1. However, it is not known how the Stack Overflow community handles obsolete answers. In this RQ, we are interested in examining how the Stack Overflow community deals with obsolete answers after such obsolescences are observed. More specifically, we would like to investigate the activities that occur after someone observes the obsolescence of an answer. Through such analysis, we expect to provide an overview of how the community handles the obsolescence of answers once they are observed and a reasonable understanding of the severity of the answer obsolescence problem for Stack Overflow developers to pay attention to.

---
[7]. www.archive.org/details/stackexchange
[8]. https://github.com/SAILResearch/replication-obsolete_answers_SO
[9]. https://stackoverflow.com/posts/comments/2293838
[10]. https://stackoverflow.com/posts/comments/65380866

**Approach:** Based on our observation during the data collection process, there are two types of actions that might occur after an answer is observed to be obsolete: 1) updating the obsolete answer (*update*); 2) creating a new updated answer (*new*). As a result of the above two types of actions, another action might occur, that is the switching of the accepted answer (*switch*). For example, the original asker may cancel the currently accepted answer and mark an updated one or a newly created one as the accepted answer. To understand what occurs after an obsolescence is observed, we perform both quantitative and qualitative analysis. An overview of the approach is presented in Fig. 3.

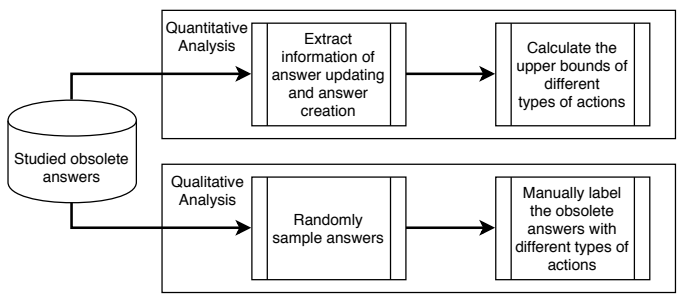

Fig. 3: An overview of our approach in RQ1.

In the quantitative analysis, we captured an overall picture about when the obsolescence is observed and how users react to obsolescence observations in terms of the three types of actions (i.e., *update*, *new*, and *switch*). To compute the number of cases in which users update the obsolete answer (type *update*), we counted the number of obsolete answers that have been edited after an obsolescence observation. Such a number gives us an upper bound estimate since updating an obsolete answer is not the only reason for editing an existing answer. We computed the number of type *new*, using a similar way as type *update*, i.e., computing an upper bound estimate. Adding updated information is one possible reason for creating a new answer, but there could be other reasons, such as adding an alternative answer. Thus, by computing the number of question threads that have new answers after an obsolescence observation, we can get an upper bound on the number of instances of type *new*. We are able to compute the number of type *switch* instances based on the historical records of answers. *However, we did not find any case of type switch. Therefore we focus the rest of our analysis on type update and new.*

In the qualitative analysis, we performed a manual study to calculate the exact occurrences of type update and new actions. We randomly sampled a statistically representative sample of 669 obsolete answers (including all their associated comments) from our studied 52,177 obsolete answers using a 99% confidence level with a 5% confidence interval. Since there are 167 (25% out of these sampled 669 answers) false positive cases, to make sure we have enough number of actual obsolete answers (to achieve a 99% confidence level with a 5% confidence interval), we kept randomly sampling from the rest of the 52,177 obsolete answers until we reach a total number of 669 actual obsolete answers. We manually performed a lightweight open coding-like process [5], [6] to check the sampled answers, their edit records, and the associated comments and other answers in the same question thread in order to label the types (update and new) of the performed actions. We also recorded the time for users to react. Note that the qualitative analysis of other RQs are also performed on these 669 actual obsolete answers.

This process involves 2 phases and is performed by the first two authors (i.e., A1–A2) of this paper:

- Phase I: A1 and A2 independently categorize the types of performed actions for each of the studied 669 answers. A1 & A2 took notes regarding the deficiency or ambiguity of the labeling for these obsolete answers.
- Phase II: A1, A2 discussed the coding results that were obtained in Phase I to resolve any disagreements until a consensus was reached. The interrater agreement of this coding process has a Cohen's kappa of 0.96 (measured before starting Phase II), which indicates that the agreement level is high [7].

### 4.1.1 Quantitative Analysis

**More than half of the studied obsolete answers were probably already obsolete as they were being posted.** Fig. 4 presents the time gap between the answer creation time and the time at which the obsolescence observation was noted. An interesting observation is that 58.4% of the studied answers were noted as obsolete within 24 hours after their creation. This suggests that more than half of the answers were probably already obsolete when they were first posted. One possible explanation is that even the answerer himself/herself did not realize that their answer is obsolete. For example, Fig. 5 shows an answerer[11] who was using an obsolete API in his original answer. A commenter pointed out within 2 minutes that the answer is obsolete, then the answerer updated his answer.

**More than half of the users do not update their answers or add new answers after their answers are noted as obsolete.** In terms of an upper bound estimation, 49.8% of the studied obsolete answers were either updated (type *update*) or added with new answers (type *new*). More specifically, less than 27.4% (upper bound) of the obsolete answers got updated after being noted as obsolete, and in 33.3% of the cases users added new answers. Note that there are answer threads that have both updated and new answers after answer obsolescence is observed in comments. We also check the editing records of the accepted answers. We observe that 44.1% of the studied obsolete answers are the accepted answers. We find that 30.7% of the obsolete accepted answers got updated (type *update*) after being

---
11. https://stackoverflow.com/questions/4650483/

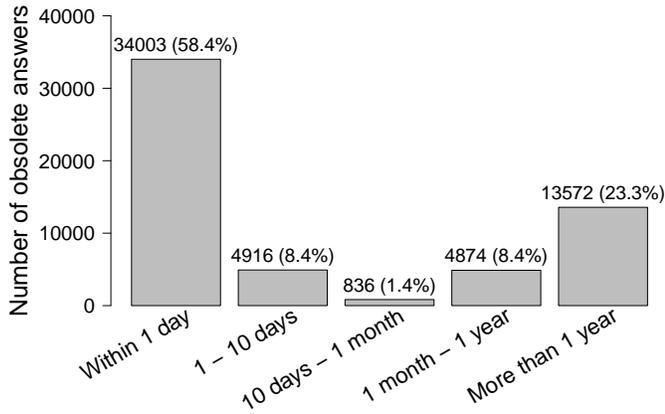

Fig. 4: Number of obsolete answers vs number of days to point out the obsolescence of the answers.

Fig. 5: An example of an answer whose poster didn't realize his answer was obsolete when he created the answer.

noted as obsolete, while only 24.8% of not-accepted answers got updated. These findings suggest that accepted answers are more likely to be updated after an obsolescence was noted compared with not-accepted answers. Nevertheless, it is interesting to note that users still do read not-accepted answers and do note their obsolescence (indicating the importance of all answers not just the accepted ones). Future studies of Stack overflow should also explore not-accepted answers instead of mostly focusing on accepted answers.

It takes 227 days on average for users to provide the first update for an obsolete answer after the obsolescence is observed in a comment, while it takes 198 days on average to add the first new answer after the obsolescence is observed.

#### 4.1.2 Qualitative Analysis

**Users updated their obsolete answers in 20.5% of the cases and added new answers in 6.3% of the cases in our qualitative study. On average, it took 118 days for users to react to an answer obsolescence observation.** For example, we present a case[12] in Fig. 6. The answer was edited on August 11, 2017 to update the obsolete answer (i.e., information about a protocol). We also notice that it took 119 days on average for users to update obsolete answers, and it took

---

12. https://stackoverflow.com/questions/3297081/

128 days on average to add new answers after an answer obsolescence was observed.

Fig. 6: An example of an obsolete answer that was updated.

> *More than half of the studied obsolete answers were probably already obsolete as they were being first posted. Most users did not update obsolete answers nor add new answers to address the obsolescence of an answer. Even for users who performed actions to deal with the obsolete answers, on average it took them 118 days after the obsolescence of the answer was noted.*

### 4.2 RQ2: Are answers to questions with particular tags more likely to become obsolete than other answers?

**Motivation:** Some particular topics (i.e., associated Stack Overflow tags) evolve more rapidly than others. For example, Android is evolving at a rather rapid pace [3]. Such rapid evolution may lead to a higher likelihood for the answers of such tags to become obsolete. Therefore, in this RQ, we examine which topics (i.e., tags in our study) of answers are more prone to have obsolete answers. By understanding this, we could provide some suggestions for the answer seekers when they search for answers on Stack Overflow (e.g., which answers relative to their associated tags require more caution since they are more likely to become obsolete). We could also provide insights into the severity of answer obsolescence across different tags, so that Stack Overflow developers could implement mechanism to solve or alleviate the specific issue.

**Approach:** We conduct a quantitative analysis to examine which tags are more likely to have obsolete answers. To understand which tags are prone to have obsolete answers, we compute the number of obsolete answers to questions that are associated with a particular tag and *normalize this number by dividing it with the total number of answers for a particular tag on Stack Overflow*.

**Results: Answers that are related to certain tags (e.g., node.js, ajax, android, and objective-c) are more likely to become obsolete.** Fig. 7 ranks the tags according to the ratio of obsolete answers to the total number of answers in each tag in our studied questions. The most obsolete-prone tag is node.js, where 0.36% of the answers with this tag have been pointed out to be obsolete. 0.34%, 0.32%, and 0.32% of the answers with tags ajax, android and objective-c are obsolete, respectively.

Due to the popularity of mobile apps, many developers are involved in mobile app development, thus leading to an increase in the number of mobile app related questions and answers on Stack Overflow. Answers related to mobile app technologies are more likely to become obsolete because of the fast progress of this field. For instance, Android has released 16 major versions and 28 levels of API from September 2008 to Aug 2018[13] and there are, on average, 115 API updates per month [3]. Another example is iOS where Apple has released 12 major versions and 103 minor versions of iOS from June 2007 to Sept 2018[14]. Such rapid updating (in both mobile operating systems and their associated tooling) makes the answers related to mobile development more likely to become obsolete. This phenomenon has also been observed by users on Meta Stack Overflow[15]. For example, a user mentions that "... Android, which as a platform is only 7 years old. It has changed drastically over that time, and answers to questions that were posed 3 or 5 years ago are out of date. In some cases the answers are inappropriate or just wrong for current developers ..."[16]. A similar situation arises to answers related to web development, such as node.js, ajax, ruby-on-rails, and jquery.

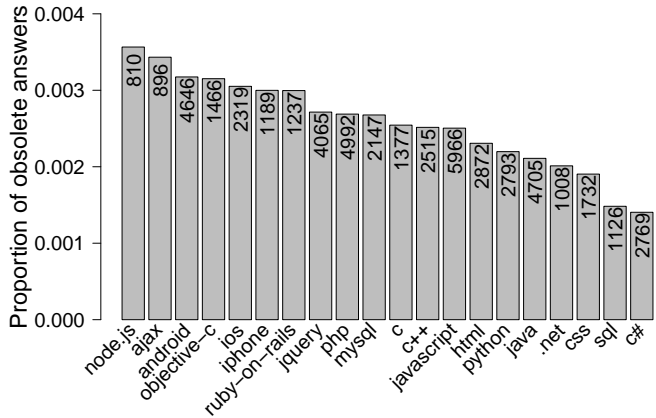

Fig. 7: The top 20 tags ranked by the ratio of obsolete answers to the total number of answers in each tag.

**There is no statistically significant difference in the obsolescence ratio (i.e., number of obsolete answers divided by total number of answers in a particular tag), between tags with large and small number of answers.** We analyze all the tags with at least 1,000 answers. The Spearman correlation between the obsolescence ratio and the number of answers in a tag is -0.049. We divide Stack Overflow communities into 7 groups based on the number of answers that are associated with a tag (i.e., 1K - 5K, 5K - 10K, 10K - 50K, 50K - 100K, 100K - 500K, 500K - 1M, and >1M), then we run the Mann-Whitney test between each pair of different groups. We also perform the Benjamini Yekutieli procedure [8] to adjust the p-values to handle the impact of multiple comparisons. We find that the adjusted p-values are greater than 0.05 for all the tests (i.e., no statistically significant difference), indicating that no matter how large the communities are, there are no differences in the obsolescence ratio of different communities. Answer obsolescence is a phenomenon across all communities on Stack Overflow.

> *Answers to questions that are associated with tags such as node.js, ajax, android, and objective-c are the most likely to become obsolete. There is no statistically significant difference in the obsolescence ratio between tags with large and small number of answers.*

### 4.3 RQ3: What are the potential reasons for answers to become obsolete?

**Motivation:** Various reasons could lead to obsolescence (e.g., a release of a new version of a framework). We are interested in investigating why answers on Stack Overflow become obsolete. Knowing this will help Stack Overflow plan better ways to avoid answer obsolescence. We can also provide insights for users to be more careful with such answers.

**Approach:** We perform a qualitative analysis to study the reasons of answer obsolescence. In this experiment, we use the same data, i.e., the randomly selected 669 answers (including all their associated comments) out of the 52,177 answers from RQ1, in order to achieve a confidence level of 99% with a confidence interval of 5%. We manually derived and categorized the obsolescence reasons (as shown in Table 1) from the randomly sampled answers threads. Note that an answer can have multiple reasons for becoming obsolete. We performed a lightweight open coding-like process [5], [6] similar to RQ1 to identify the reasons of obsolescence. This process involves 3 phases and is performed by the first two authors (i.e., A1–A2) in this paper:

- Phase I: A1 derived a draft list of obsolescence reasons based on 50 random answers. Then, A1 and A2 use the draft list to categorize the answers collaboratively. During this phase the reasons were revised and refined.
- Phase II: A1 and A2 independently applied the resulting reasons from Phase I to categorize all 669 answers. A1 & A2 took notes regarding the deficiency or ambiguity of the labeling for obsolete answers. During this phase no new labels (i.e., reasons) were introduced.
- Phase III: A1, A2 discussed the coding results that were obtained in Phase II to resolve any disagreements until a consensus was reached. The inter-rater agreement of this coding process has a Cohen's kappa of 0.76 (measured before starting Phase III), which indicates that the agreement level is substantial [7].

During our manual study process, we also labeled whether the obsolescence is a legacy or invalid obsolescence (see Section 2).

---

13. http://socialcompare.com/en/comparison/android-versions-comparison
14. https://en.wikipedia.org/wiki/IOS_version_history
15. https://meta.stackoverflow.com/
16. https://meta.stackoverflow.com/questions/309152/

TABLE 1: Reasons for obsolescence

| Reason | Definition | Example |
|---|---|---|
| Third Party Library | An answer becomes obsolete due to third party libraries, Application Programming Interfaces (APIs), or frameworks becoming obsolete. | A comment points out that the way to delete a project in Google APIs Console has become obsolete[17]. |
| Programming Language | Answer obsolescence is caused by obsolete features of the programming language and/or its standard APIs. | A comment points out that the -client option is ignored by a 64-bit capable JDK since Java 6[18]. |
| Reference | References in an answer are obsolete. | A comment points out that the link to a whitepaper with detailed benchmarking for the Oracle TimesTen in-memory database is dead[19]. |
| Tool | Tool information is obsolete, such as an old version. | A comment points out that a solution is out of date for Microsoft Kinect SDK version 1.0[20]. |
| Mobile OS | An answer becomes obsolete due to an obsolete mobile platform. | A comment points out the event handling syntax for Mono for Android 4.2 is out of date[21]. |
| Non-mobile OS | An answer becomes obsolete due to an obsolete non-mobile OS platform. | A comment points out that in order to work on macOS Sierra instead of macOS El Capitan, the new option is `--install-dir /usr/local/bin`[22]. |
| Protocol | An answer is obsolete because a protocol is updated. | A comment points out that the internet text messages RFC 822 was replaced by RFC 2822[23]. |

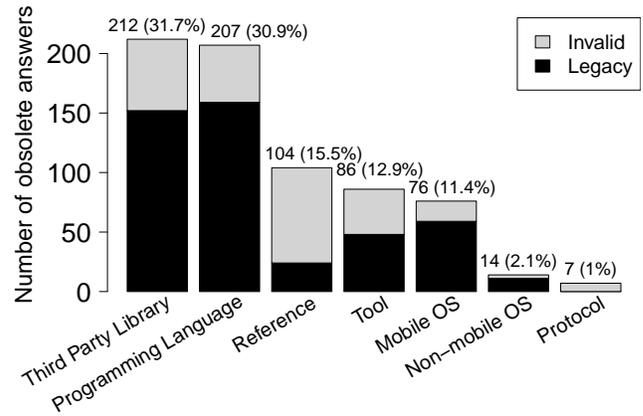

Fig. 8: Number and percentage of each obsolescence reason based on our manual analysis. The figure also shows the proportion of legacy (black) and invalid (gray) obsolescence.

**Results: 31.7% of the studied answers (after removing false positives) became obsolete due to the evolution of their associated third party libraries.** The number of occurrence and percentage of each obsolescence reason is shown in Fig. 8, as well as the proportion of legacy or invalid obsolescence for each obsolescence reason. In our qualitative study, we find that most answers became obsolete due to the evolution of their associated third party libraries. In addition, **30.9% of the studied answers became obsolete due to the evolution of their programming languages**. Stack Overflow covers a broad range of questions and answers across various programming languages and third party libraries, and it is very common for programming languages/third party libraries to release new versions, thereby making the older versions possibly obsolete. For example, in a question of how to serialize and restore an unknown class in c#, an answer[24] suggested to use SoapFormatter instead of XmlSerializer. Another user posted a comment 3 minutes later stating that "this class is obsolete. Use BinaryFormatter instead", including the .NET Framework version number and a reference link. Based on this observation, we recommend that users provide a version number for their answers, then Stack Overflow can note the active versions when an answer was posted and note in the UI how many versions come after it.

---

17. https://stackoverflow.com/posts/comments/56423259/
18. https://stackoverflow.com/posts/comments/59707599/
19. https://stackoverflow.com/posts/comments/803108/
20. https://stackoverflow.com/posts/comments/12009382/
21. https://stackoverflow.com/posts/comments/14581496/
22. https://stackoverflow.com/posts/comments/75888652/
23. https://stackoverflow.com/posts/comments/14278476/
24. https://stackoverflow.com/questions/590722/

**15.5% of the answers are obsolete due to obsolete references. 11.9% of the 5.5 million links (that are mentioned on Stack Overflow answers) are no longer available.** Obsolete references include URL links, cited books, videos, and so on. Although it is convenient for a user to post an answer simply by referring to external URLs, it is common for references to become obsolete because the source of the reference may not be well maintained over time. This is especially a problem when users write an answer without providing too much concrete content, but instead simply offering URLs as the solution. In total, there are 5.5 million links in the 7.3 million answers on Stack Overflow. To better understand the obsolete URLs on Stack Overflow, we check all 5.5 million links to verify if they are still accessible (i.e., by returning 200 status code when requesting the URL). As of September 2018, we find that 11.9% of these links are no longer accessible.

**12.9% of the studied obsolete answers are due to outdated tools, and 27.9% of these outdated tools are related to IDEs.** To further understand what types of tools are more likely to be associated with obsolete answers, we manually study the related answer threads. Among these tools, 27.9% are related to IDEs, such as Visual Studio, Eclipse, Xcode, and Android Studio. For example, in an outdated answer for Xcode, the commenter not only pointed out the obsolescence, but also provided an updated answer[25]. One possible explanation is that IDEs are frequently updated in order to provide support for evolving programming languages and environments (e.g., mobile development).

Besides these obsolescence reasons, we also observe others, such as obsolete operating systems, and protocols. For example, a comment[26] in an answer pointed out that since Windows 7 cacls is deprecated for displaying and modifying access control lists (ACLs).

Obsolete answers should not simply be removed as a solution because they may still be applicable to users who are using legacy technologies/systems. We find that 63.8% of the studied obsolete answers belong to the legacy category. **However, we observe that the studied answers that**

---

25. https://stackoverflow.com/posts/comments/16320934/
26. https://stackoverflow.com/posts/comments/54010530/

**are related to protocols are all invalid.** This is reasonable since once a protocol becomes obsolete, it is most likely no longer used anymore. We get the complete list of RFCs[27] as of May 2018. This list contains the 8,286 RFCs, in which 1,188 RFCs are obsolete because of 1,112 newly added RFCs. We collected all answers (i.e., 21,591) containing "RFC" information from Stack Overflow, and we find that the RFCs in 10,793 answers became obsolete (i.e., were replaced by new RFCs). However, among such obsolete answers, only 611 answers were updated to reflect the new RFC versions. In other word, **only 5.7% of answers mentioning obsolete RFCs were updated to reflect the new RFC version.**

> *The majority of answers are obsolete due to the evolution of their associated third party libraries, programming languages, and references. Therefore, users need to pay more attention to such answers when looking for answers on Stack Overflow.*

### 4.4 RQ4: Who observes obsolete answers and what evidence do these observers provide?

**Motivation:** Uncovering obsolete knowledge on Stack Overflow is not trivial, especially if the user is not an expert in the specific knowledge domain. Therefore, it is essential to identify experts who might observe answer obsolescence and support their observations. In this RQ, we examine who identifies obsolete answers. Furthermore, we are interested in investigating how they support their obsolescence observation. By analyzing these aspects, we expect to get insights into how to assist users on Stack Overflow to identify obsolete answers.

**Approach:** To understand who observes the obsolescence of an answer, we first perform a quantitative study on all the studied answer threads. Based on the role of the user who notes the obsolescence observation in an answer thread, we categorize observers into one of the following 5 groups:

1) **Asker**: the user who posted the question;
2) **Answerer**: the user who posted the obsolete answer;
3) **Other answerer**: the user who posted another answer other than the obsolete one;
4) **Commenter**: the user who posted comments in the question thread;
5) **Outsider**: the user who never had any prior activities (including posting question, answer or comment) in that question thread.

We refer to an asker, answerer, other answerer(s), or commenter who are involved in the question thread (groups 1 – 4) as an *insider* (since they were involved earlier in the question thread).

To understand the type of evidence that users provide when noting the obsolescence of an answer, we performed a qualitative study. We used the studied answers from RQ1. We manually extracted and categorized the evidence of obsolescence from the sampled answers. We performed a lightweight open coding-like process [5], [6] as mentioned in RQ3. We categorized the support evidence for obsolete answers into 8 types, as shown in Table 2. The inter-rater agreement of this coding process has a Cohen's kappa of 0.95, which indicates that the agreement level is high [7].

[27]. https://www.ietf.org/download/rfc-index.txt

TABLE 2: Types of support evidence for an obsolescence observation.

| Type | Definition |
| --- | --- |
| Provide updated info | The user provides updated information as an explanation why an answer is obsolete. |
| Provide version info | The user mentions the version number of either the obsolete answer (e.g., framework) or the updated information. |
| No support | No supportive material is given to prove the answer is obsolete. The user simply claims that something is obsolete. |
| Provide links | The user posts a link as a further reference to her/his obsolescence observation. |
| Highlight time | The user mentions the time when the answer worked. |
| Provide running errors | The user shows the running errors due to the obsolescence. |
| Refer to other answers | The user points to another answer on Stack Overflow to support why the current answer is obsolete. |
| Refer to this answer | The user points to this answer because it updated the obsolete content. |

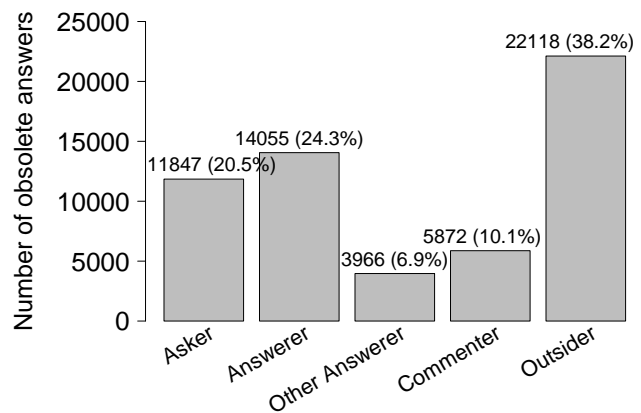

Fig. 9: The number (as well as the percentage) of the obsolete answers that are observed by each type of user. A role of user is assigned using the following priority: asker > answerer > other answerer > commenter > other user. For example, if a user has multiple roles, such as an answerer and a commenter, we consider the user as an answerer.

**Results: The obsolescences of answers are more frequently observed by outsiders (38.2%), compared to askers (20.5%) and answerers (24.3%)** The number and proportion of obsolete answers that were observed by each group of users (i.e., asker, answerer, other answerer, commenter, and outsider) are shown in Fig. 9. Only 24.3% of the obsolete answers were observed by answerers. 10.1% of the obsolete answers were observed by commenters. 6.9% of the obsolete answers were observed by other answerers in the same question thread. 20.5% of the obsolete answers were observed by askers. The lowest proportion among the insiders are other answerers. The rest of the obsolete answers (38.2%) are observed by users who have never participated in the discussion before observing that the answer is obsolete.

In summary, only 24.3% of the obsolete answers were observed by answerers. One possible reason is that some answerers are no longer active on Stack Overflow. Another possible reason is that even if the answerers are still active on Stack Overflow, they may not really want to maintain their answers after a long period of time. Even worse, they may not even be active in that domain anymore. For example, one user asked how to handle obsolete answers, and one commenter mentioned that *"Two years down the line I don't want to have to regularly rework my answers. I might not even be active in that field anymore"*[28]. Therefore, it's very important for Stack Overflow to encourage the whole community, not just the answerers to maintain answers by taking care of obsolete answers.

askers are more likely to report runtime errors. One possible explanation is that askers are more likely to have a chance to run the code that is proposed in the answer and find out that it does not work due to runtime errors. Then, they report the error in the comment. In general, outsiders are the main evidence providers for pointing out obsolete answers.

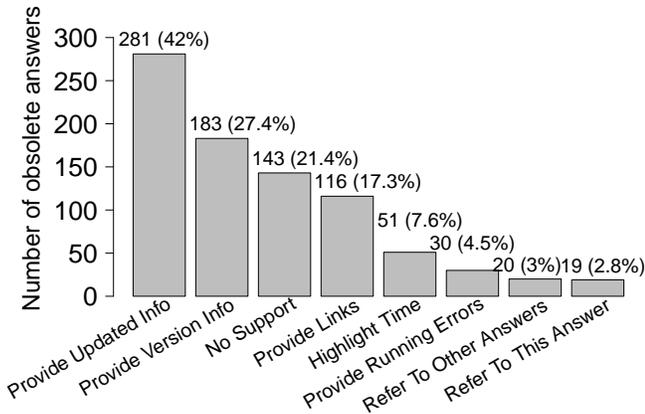

Fig. 10: The proportion of each type of evidence that users provide when pointing out obsolescence.

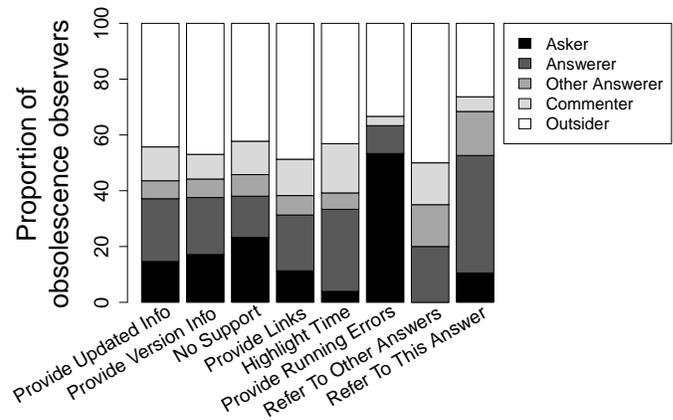

Fig. 11: The proportion of each type of evidence that different observers provide when pointing out the obsolescence of an answer.

**The majority (78.6%) of the obsolete observations are supported with evidence (e.g., updated information, a version information, or a reference).** Fig. 10 shows the proportion of each type of supporting evidence for obsolescence observations. An obsolescence observation could have multiple types of support evidence. For example, a user can provide both version information and a link to the new version. We observe that in the majority of cases, users provide supporting evidence (e.g., updated information and a version information). In 42% of the cases, users provide updated information about the obsolete answers. For example, in a comment[29], the user not only pointed out that numpy is out of date, but also provided the code to check the numpy version in the code to install the latest version. Such cases are not rare; we observe that 44.8% of cases a solution (an updated answer) is provided in the comments. Furthermore, version numbers are also used by some users to support obsolescence observation. Once a version number is given, it is convenient for users to identify the obsolete knowledge. We find that 27.4% of obsolescence observations mentioned version numbers. For example, in an answer that uses AutoMapper (a convention-based object-to-object mapper and transformer for .NET), one comment[30] started with *"as of AutoMapper 4.2 Mapper.CreateMap() is now obsolete ..."*. However, we find that 21.4% of obsolescence observations do not provide any supporting evidence. During our qualitative study, we find other types of support for obsolescence observations. For example, 7.6% of obsolescence observations are supported by highlighting time information (e.g., the validity period for the answer) related to the obsolescence.

Obsolescence observers tend to provide different evidence to support their observations. As shown in Fig. 11,

*The majority of the obsolete answers were not observed by the original answerers. To help resolve obsolete answers, Stack Overflow should develop mechanisms to encourage the various members of the Stack Overflow community to maintain and flag obsolete answers. We also find that most (78.6%) obsolescence observations are supported by evidence.*

## 5 IMPLICATIONS AND ACTIONABLE SUGGESTIONS FROM OUR FINDINGS

### 5.1 Actionable Suggestions for Stack Overflow

**An automated tool could be built to identify existing obsolete answers on Stack Overflow, or help answerers identify obsolete answers in real-time during answer creation.** We find that more than half of the obsolete answers were identified as obsolete within 24 hours of their initial posting, which indicates that users may not even realize that their posted answers are already obsolete. An automated tool could be developed to identify the possible obsolescence of an answer as it is being typed in. More specifically, we observed that there are many obsolescence reasons and the two major ones are related to third party libraries and programming languages. Future research could possibly leverage the evolution information of third party libraries and programming languages to detect the obsolescence of related answers. For example, a tool could analyze third party libraries to check their latest version, or the time of their latest update, and determine the valid API version for an API related answer so that version information is highlighted in appropriate answers. As an example, Tran et al. [9] automatically detected outdated information on Wikipedia by using pattern-based fact extraction from both Wikipedia and the web. A similar tool may be developed

---

28. https://meta.stackexchange.com/posts/comments/21537/
29. https://stackoverflow.com/posts/comments/56525745/
30. https://stackoverflow.com/posts/comments/58514542/

to scan existing answers and label those that are obsolete with valid version information of a library or programming language where applicable.

**An automated mechanism to detect obsolete references is needed. We scanned all links (i.e., 5.5 million) in Stack Overflow answers and found that as of September 2018, 11.9% of the links are inaccessible.** Hence, Stack Overflow could scan links to identify the availability of links. Similar to the dead link template and other inline cleanup tags (such as obsolete source) on Wikipedia[31], Stack Overflow could also include a "dead link" tag as well as the last retrieved time once an obsolete link is detected. As a result, users are made aware of obsolete links when reading the answer, and users who posted obsolete links could also be notified when their links are detected as obsolete. Additional actions are therefore encouraged, such as updating obsolete links or archiving snapshots of links as soon as they are created.

**Our heuristic-based approach for identifying obsolete answers using comments has an accuracy of 75%. Future work could improve the accuracy of our approach using machine learning techniques (e.g., classification).** Machine learning techniques could be applied to identify whether a comment indicates that an answer is obsolete based on the content of the comment and other features, such as the associated tags of the answer, and answer/comment score. Note that we characterize the false positives in the data collected by our heuristic-based approach (in Section 3), so future work could pay special attention to these corner cases in order to improve the accuracy of any automated approach. For example, a comment mentioning "function ABC was replaced by XYZ in year N" would be a strong indication of an obsolete answer. As a result, such comments could be highlighted to assist users in identifying obsolete answers.

**Stack Overflow should develop mechanisms to encourage users (especially question thread insiders) to pay more attention to the obsolescence of answers (their own or others') and make efforts to maintain any obsolete answers.** In RQ1, we find that only around 1 out of 4 users updated their answers when their answers were noted as obsolete. Moreover, it took users about 4 months on average (i.e., 118 days) to update their answers or add new updated answers. In other words, users do not pay much attention to the obsolescence of their answers and do not frequently maintain their answers. For example, a comment of an obsolete answer mentioned that the answer was obsolete and asked the answerer to update it. The answerer replied in comment "*Feel free to update the answer yourself, if you like. I honestly would, but I don't have the time*"[32]. The gamification system (e.g., badges and/or reputation scores) should be adjusted to encourage users to identify and update obsolete answers. For example, Stack Overflow could reward badges or reputation scores to users who identify or maintain obsolete answers.

### 5.2 Actionable Suggestions for Users

**Answerers are encouraged to include relevant information about the valid version or the time of their knowledge when creating answers.** In RQ4, we observe that 78.6% of the obsolescence observations included supportive evidence, such as when the answer became obsolete (e.g., time and version information). Such information is very helpful for answer seekers to verify whether the knowledge in the answers is still valid or not (especially for their context).

**Answer seekers are encouraged to carefully go through the comments that are associated with answers in case these answers become obsolete, especially for answers that are related to web and mobile development, such as node.js, ajax, android, and objective-c.** In RQ2, we observe that answers related to some specific tags are more likely to become obsolete, such as tags that are related to mobile development (e.g., Android and iOS) and web development (e.g., node.js and ajax). Therefore, answer seekers are encouraged to pay more attention when reading through answers that are related to such tags. One actionable way is to go through the comments under accepted answers or not-accepted (yet highly voted) answers, which may have useful information to indicate whether the answer has became obsolete or not. Even more, in 44.8% of the observed obsolete answers, a comment provided a solution to update the answer. In addition, we strongly advise users to carefully read all highly ranked comments when reading an answer, since we observe that 73.5% of the comments that indicate the obsolescence of an answer are the top 1 ranked comment for the obsolete answers.

### 5.3 Feedback from Stack Overflow

To understand whether our research uncovered a relevant problem on Stack Overflow and whether our findings are useful for Stack Overflow, we shared our findings with members of the Stack Overflow team. They concurred with our findings and mentioned that it is interesting to see a breakdown of this problem (*"obsolete info is an ongoing issue on the site, so it's interesting to see this breakdown of how that issue manifests itself"*). They asked us to examine whether the answer obsolescence issue would vary based on different community sizes. We found that answer obsolescence is a widespread issue that is not influenced by the size of the tag (the details of this analysis is included in RQ2). Moreover, they were specifically interested in our analysis about the version information of platforms and programming languages. Based on our findings, the Stack Overflow team was also interested in investigating approaches to generate tags that indicate the valid version for a framework, an API, or a programming language for an answer. Future research efforts should continue working with the Stack Overflow team to solve/alleviate the obsolete problem.

## 6 THREATS TO VALIDITY

**External validity:** Threats to external validity are related to the generalizability of our findings. In this study, we focus on Stack Overflow, which is one of the most popular and largest Q&A websites for developers; hence, our results may not generalize to other Q&A websites. To alleviate this threat, more Q&A websites should be studied in the future. We needed to conduct several qualitative analysis in our RQs; however, it is impossible to manually study all answers. To minimize the bias when conducting our qualitative analysis, we took statistically representative random samples of all relevant revisions, in order to ensure a 99% confidence level and 5% confidence interval for our observations [10].

---

31. https://en.wikipedia.org/wiki/Template:Dead_link
32. https://stackoverflow.com/posts/comments/61093395/

**Internal validity:** Threats to internal validity are related to experimenter bias and errors. Our study involved qualitative analysis in RQs. To reduce the bias, each answer was labeled by two of the authors and discrepancies were discussed until a consensus was reached. We also showed that the level of inter-rater agreement of the qualitative studies is high (i.e., the values of Cohen's kappa ranged from 0.76 to 0.96). Another threat to our study is related to our data collection process. Due to the large number of answers and lack of mechanism on Stack Overflow to identify obsolete answers, we used a heuristic-based approach to uncover obsolete answers. The accuracy of our heuristic-based approach is 75% based on our manual verification, which implies that there may be noise in our quantitative study. Hence we followed all presented quantitative studies with qualitative studies of randomly representative samples. Future study should develop a more accurate method to identify the obsolescence of an answer on Stack Overflow. In RQ1, a quantitative analysis shows an upper bound for both the proportion of obsolete answers that were updated and the proportion of new answers that were created after such obsolete answers. The values do not show how many answers are actually updated or created due to answer obsolescence, but only indicate an upper bound of such cases. Other reasons (e.g., provide alternative solutions) could cause users to update and/or add answers. This represents a possible threat to the internal validity of this particular analysis. To tackle this concern, we performed a qualitative study in RQ1 to manually analyze how many answers are updated or added due to answer obsolescence. An additional threat lies in the evaluation of our heuristic approach to find obsolete answers. The first two authors of the paper evaluated this heuristic approach. We calculated Cohen's kappa to measure the inter-rater agreement between both authors and the agreement is high (i.e., 0.76).

# 7 RELATED WORK

We compare our study with prior studies as shown in Table 3.

## 7.1 Understanding and Improving the Quality of Posts On Stack Overflow

One significant challenge that Q&A websites have is ensuring the quality of their knowledge [21]. Therefore, numerous studies have been done to better understand and improve the quality of knowledge on Q&A websites. The majority of prior studies define the quality of content on Stack Overflow from the presentation aspect (e.g., code and text) [22]–[30]. For example, Asaduzzaman et al. studied unanswered questions on Stack Overflow and revealed reasons for such unanswered questions [22]. Zhang et al. conducted an empirical study on the prevalence and severity of API misuse on Stack Overflow [28]. Chen et al. proposed a deep learning approach to help users on Stack Overflow fix grammar issues based on prior editing records [30]. Wang et al. analyzed how the badge system impacts answer revision on Stack Overflow, and found that the current system fails to consider the quality of revisions [31].

Some studies also consider the quality of content in terms of the time aspect; namely, obsolescence. Wu et al. surveyed 453 users on Stack Overflow and found that outdated code is one of their major complaints [4]. Ragkhitwetsagul et al. studied the answer obsolescence issue on Stack Overflow by conducting online surveys [11]. They found that half of the top answerers in their survey are aware of obsolete code snippets. However, participants rarely or never fix obsolete code snippets. Ragkhitwetsagul et al. also analyzed Java code snippets that were copied to Stack Overflow [12], and found that 66% of such code snippets are outdated. Fischer et al. noted outdated SSL/TLS versions and outdated algorithms when they analyzed security-related code snippets in Android-related posts on Stack Overflow [13].

We study the answer obsolescence issue across all domains on Stack Overflow instead of focusing on specific domains (e.g., code snippets), in order to provide insights for Stack Overflow to alleviate the general answer obsolescence problem and improve the overall quality of Stack Overflow answers.

## 7.2 API Obsolescence in Software Engineering

Obsolescence is a common issue for software systems. Technology consulting firms estimate that 180-200 billion lines of legacy code is still in active use [32]. One reason for obsolescence is that the used APIs become obsolete due to deprecation. A significant amount of studies have examined API deprecation [3], [14], [15], [33]–[36]. For example, Beyer et al. categorized Stack Overflow questions related to Android into 7 types, including API change [33]. Linares-Vásquez et al. studied how developers react to Android APIs deprecation on Stack Overflow [14]. McDonnell et al. studied how APIs evolved in the Android ecosystem and found that 28% of API calls are outdated with a 16 months lag time (i.e., the time between commit and the API release) [3]. Zhou et al. proposed an approach to detect deprecated Android API usages in source code examples on Stack Overflow [34]. Reboucas et al. noted the API obsolescence issues are often due to the rapid development cycles in the Swift programming language [35].

Different from prior studies, which only focused on APIs, we focus on the general obsolescence of all answers on Stack Overflow and investigate the characteristics of such obsolete answers. We also study how users deal with obsolete answers on Stack Overflow and provide actionable suggestions for Stack Overflow.

## 7.3 Leveraging the Knowledge from Stack Overflow

Stack Overflow accumulates a large amount of knowledge and researchers have done a remarkable number of studies to leverage the knowledge on Stack Overflow to facilitate development and maintenance activities [17]–[20], [37]–[39]. For example, Zagalsky et al. recommended high-quality code by leveraging knowledge from Stack Overflow [17]. Treude et al. developed a tool to enrich API documentation with "insight sentences" extracted from Stack Overflow [18].

TABLE 3: Comparison between our findings and prior studies.

| Topic | Prior studies | Our study |
| --- | --- | --- |
| Understanding and improving the quality of posts on Stack Overflow | Prior studies noted the existence of outdated code snippets using user surveys [4] [11] [12]. Prior studies analyzed Java code in accepted answers [12], or security-related code in Android posts [13]. | We study the answer obsolescence issue across all domains on Stack Overflow instead of focusing on specific domains (e.g., code snippets), in order to provide insights for Stack Overflow to alleviate the answer obsolescence problem and improve the long-term quality of Stack Overflow answers. We also analyze how Stack Overflow users deal with obsolete answers, i.e., they rarely maintain obsolete answers. |
| API obsolescence in software engineering | Prior studies analyzed how obsolete APIs impact the software engineering ecosystems, such as Stack Overflow [14], and Smalltalk projects [15]. Prior studies also investigated how APIs evolved in the Android ecosystem [3] and in API documentation [16]. | Instead of focusing only on obsolete APIs, especially Android APIs, we find that more than half of the obsolete answers are due to other reasons, such as programming language, references, and tools. We also study how users deal with obsolete answers on Stack Overflow and provide actionable suggestions for Stack Overflow. |
| Leveraging the knowledge from Stack Overflow | Prior studies leveraged Stack Overflow posts to enhance existing software artifacts, such as source code [17], API documentation [18], JavaDoc [19], and source code comments [20]. | We highlight the potential risk of answer obsolescence on Stack Overflow. We provide actionable suggestions for both Stack Overflow and its users (including answerers and answer seekers) to manage, identify and avoid obsolete answers. For example, we provide actionable suggestions towards building automated tools to detect obsolete answers on Stack Overflow. |

Vassallo et al. extracted discussions from Stack Overflow to generate JavaDoc automatically [19]. Wong et al. leveraged questions and answers on Stack Overflow to automatically generate comments in system source code [20]. Gao et al. proposed an automated approach to fix recurring crash bugs by leveraging information (e.g., questions with similar crash traces) on Q&A websites [38]. Wang et al. leveraged the tag information on Stack Overflow to infer semantically related software terms [39].

Instead of leveraging the knowledge from Stack Overflow, we study the knowledge obsolescence on Stack Overflow. Our finding indicates that many answers on Stack Overflow may become obsolete, which may affect the quality of the content that is produced by the above-mentioned techniques. Therefore, further research should take caution when leveraging the knowledge from Stack Overflow.

## 8 Conclusion

In this paper, we present an empirical study of the obsolete knowledge on Stack Overflow, as an inevitable step towards understanding the evolution of knowledge on Stack Overflow. We find that: 1) Answers in certain tags (e.g., node.js, ajax, android, and objective-c) are more likely to become obsolete mainly due to the evolution of their associated third party libraries and programming languages. 2) Most of the studied obsolete answers are pointed out by non-answerers and are supported by evidence. 3) When an obsolete answer is identified, only a small proportion of such answers are updated afterwards. More importantly, more than half of the obsolete answers were probably already obsolete when they were posted. Based on our findings, we offer the following suggestions: 1) Stack Overflow should develop mechanisms (i.e., incentive systems) to encourage the whole community to identify and/or maintain obsolete answers. 2) Answerers are encouraged to include information of the valid version or time of the knowledge when creating answers. 3) Answer seekers are encouraged to go through all the information in an answer thread carefully in case someone had pointed out the obsolescence of an answer, especially for the answers that are related to web and mobile development.

There are two possible directions for future work. First, we encourage future studies to develop advanced approaches to detect obsolete knowledge on Stack Overflow. For example, machine learning techniques can be employed to detect the comments that indicate obsolescence based on the semantic meaning of the text instead of keywords matching. Second, we encourage future studies to develop approaches to extract useful information from the comments so that answer seekers could easily find the useful information from the list of long and unorganized comments.

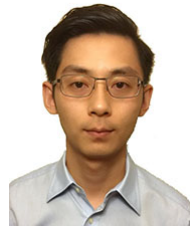
**Haoxiang Zhang** Haoxiang Zhang is currently working toward a PhD degree in the School of Computing at Queen's University, Canada. He received his PhD degree in Physics from Lehigh University, Bethlehem, Pennsylvania in 2013. His research interests include machine learning in software analytics, empirical software engineering, and mining software repositories. More information at: https://haoxianghz.github.io/.

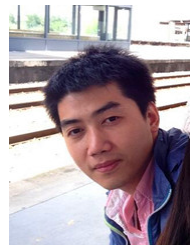
**Shaowei Wang** Shaowei Wang is a postdoctoral fellow in the Software Analysis and Intelligence Lab (SAIL) at Queen's University, Canada. He obtained his PhD from Singapore Management University, and BSc from Zhejiang University. His research interests include code mining and recommendation, software maintenance, developer forum analysis, and mining software repositories. More information at: https://sites.google.com/site/wswshaoweiwang/.

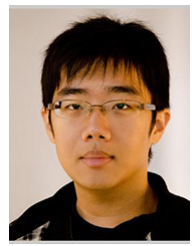
**Tse-Hsun (Peter) Chen** Tse-Hsun (Peter) Chen is an Assistant Professor in the Department of Computer Science and Software Engineering at Concordia University, Montreal, Canada. He obtained his BSc from the University of British Columbia, and MSc and PhD from Queen's University. Besides his academic career, Dr. Chen also worked as a software performance engineer at BlackBerry for over four years. His research interests include performance engineering, database performance, program analysis, log analysis, and mining software repositories. Early tools developed by Dr. Chen were integrated into industrial practice for ensuring the quality of large-scale enterprise systems. More information at: http://petertsehsun.github.io/.

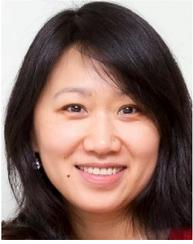

**Ying Zou** Ying Zou is the Canada Research Chair in Software Evolution. She is a professor in the Department of Electrical and Computer Engineering, and cross-appointed to the School of Computing at Queens University in Canada. She is a visiting scientist of IBM Centers for Advanced Studies, IBM Canada. Her research interests include software engineering, software reengineering, software reverse engineering, software maintenance, and service-oriented architecture. More about Ying and her work is available online at: http://post.queensu.ca/~zouy/.

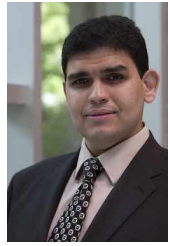

**Ahmed E. Hassan** Ahmed E. Hassan is a Canada Research Chair in Software Analytics and the NSERC/Blackberry Industrial Research Chair with the School of Computing, Queen's University, Kingston, ON, Canada. His industrial experience includes helping architect the Blackberry wireless platform, and working for IBM Research at the Almaden Research Lab and the Computer Research Lab at Nortel Networks. Early tools and techniques developed by his team are already integrated into products used by millions of users worldwide. He is the named inventor of patents at several jurisdictions around the world including the United States, Europe, India, Canada, and Japan. Dr. Hassan serves on the editorial board of the IEEE Transactions on Software Engineering, the Journal of Empirical Software Engineering, and PeerJ Computer Science. He spearheaded the organization and creation of the Mining Software Repositories (MSR) conference and its research community. More about Ahmed and his work is available online at: http://research.cs.queensu.ca/~ahmed/.